\documentclass[12pt]{article}
\usepackage{sw20bams}

\input tcilatex
\QQQ{Language}{
American English
}

\begin{document}

\title{On Mathematical Structure of Effective Observables}
\author{C. P. Viazminsky$^a$ and James P. Vary$^b$ \\
a) Department of Physics, University of Aleppo, Syria\\
and International Institute of Theoretical and Applied Physics,\\
Iowa State University, Ames, Iowa 50011, USA\\
b) International Institute of Theoretical and Applied Physics\\
and Department of Physics and Astronomy,\\
Iowa State University, Ames, Iowa 50011, USA}
\maketitle

\begin{abstract}
We decompose the Hilbert space of wave functions into two subspaces, and
assign to a given observable two effective representatives that act in the
model space. The first serves to determine some of the eigenvalues of the
full observable, while the second serves to determine its matrix elements,
in any basis in one of the subspaces, in terms of quantities pertaining to
the model space. We also show that if the Hamiltonian of a physical system
possesses symmetries then these symmetries continue to hold for its
effective representatives of the first type. Maximum information about the
system can be obtained in terms of two sets of effective representatives.
The first set of representatives is complete. Other observables that do not
commute with all members of the complete set have only one type of
representative.
\end{abstract}

\section{\ Introduction}

Effective operators are often used in nuclear, atomic and molecular physics.
The general scheme aims to construct from the Hamiltonian of the system,
acting on the Hilbert space of wave functions, an operator that acts on a
low dimensional space, so that the eigenvalues of the latter operator are
also eigenvalues of the full Hamiltonian of the given system [1-9]. The low
dimensional space, we have mentioned, is called a model space and the
operator acting on it to produce some of the eigenvalues of the full
Hamiltonian is called an effective Hamiltonian, or an effective
representative of the Hamiltonian. The latter requirement does not determine
an effective representative uniquely. A general class of effective
representatives was obtained by Suzuki \cite{Suzuki} who also delineated
forms according to the role of an arbitrary parameter, the starting energy,
in the iterative method of solution.\cite{Kuo}, or according to their
Hermiticity. Hermitian forms have been introduced or adopted by many
researchers [10-17]. A standard non-Hermitian form \cite
{Brandow,fields,suzu??} is relatively simple, and is commonly used for
implementing the scheme of effective representatives.

Our present work, which is concerned with the effective representation of
any observable in the standard non-Hermitian scheme, has the following
objectives:

1. To establish the equivalence between the decoupling condition on the
transformed observable and a corresponding condition on its transformed
eigenfunctions.

2. To show that the decoupling equation always has solutions and to specify
the maximum number of inequivalent solutions.

3. Starting from a complete set of observables associated with the physical
system, we construct a complete set of effective representatives, and prove
accordingly that the symmetries of the Hamiltonian are carried over to the
effective representatives.

4. Two effective representatives can be constructed associated with every
observable. The first representative corresponds to the standard
non-Hermitian form and gives some of the eigenvalues of the original
observable. The second representative is Hermitian and has the property that
the matrix elements of the original observable, in any basis of the subspace
that is mapped onto the model sace, can be calculated in terms of this
representative and the projected basis in the model space.

\section{The Model Space}

The truncated Hilbert space of square integrable functions associated with
the system, denoted by $H_N$, consists of all $N-$columns with complex
entries. $H_N$ is just the unitary space of complex numbers $C^N$ through
the isomorphism $\psi \in H_N\longleftrightarrow \psi ^t\in C^N,$ where $(t)$
denotes the transpose. The standard basis in $H_N$ will be denoted by $%
e_i(i=1,....,N)$, so that 
\begin{equation}
e_{1=}(1,0,....,0)^t,\;e_{2=}(0,1,....,0)^t,.........,e_{_N=}(0,....,0,1)^t
\label{e1}
\end{equation}
\quad \quad

Let $K$ be a distinct subset of $d$ elements of the set $\{1,2,....,N\}.$
The subspace generated by the subset of basis elements $\{e_k:k\in K\}$ will
be denoted by $\Pi _K,$ and will be called a model space. The projection on $%
\Pi _K$ will be denoted by $P_K$, whereas $Q_K$ will denote the projection
on the orthogonal complement $\Pi _K^{\bot }=H_N\ominus \Pi _K$. It follows
that $P_K+Q_K=I,\,P_KQ_K=Q_KP_K=0.\,$\thinspace If it is desired, one may
rearrange the order of the basis elements (\ref{e1}) so that the vectors $%
e_i(i\in K)$ are placed first. We shall assume that such reordering is done
whenever it is necessary, and drop the index $K$, if no ambiguity arises.
The symbol $P$ accordingly, will denote a projection on some model space $%
\Pi $. The reordering operation is particularly useful when we have to
represent vectors and operators in matrix form.

Let $S$ be an operator in $H_N$ such that 
\begin{equation}
S=QSP  \label{e2}
\end{equation}
It follows that $S^2=0,$and hence $e^{\pm S}=1\pm S.$ Equation (\ref{e2})
implies also that

\begin{equation}
S=\left( 
\begin{array}{cc}
0_d & 0 \\ 
s & 0
\end{array}
\right)  \label{e3}
\end{equation}
where $0_d$ is the nil $d\times d$ matrix. Consider the transformation 
\begin{equation}
e^{-S}:H_N\rightarrow H_N\,,\,\,\,\,\,\psi \rightarrow \widetilde{\psi }%
=(1-S)\psi .  \label{e4}
\end{equation}
Setting $\psi =\tbinom \alpha \beta $ where $\alpha ^t\in C^d$, we write 
\begin{equation}
\widetilde{\psi }\equiv \tbinom{\widetilde{\alpha }}{\widetilde{\beta }}%
=\left( 
\begin{array}{cc}
1 & 0 \\ 
-s & 1
\end{array}
\right) \binom \alpha \beta =\binom \alpha {\beta -s\alpha }  \label{e5}
\end{equation}
It is apparent that the mapping $e^{-S}\,$is determined by $S$ given by (\ref
{e3}), which in turn is determined by $s:\Pi \rightarrow \Pi ^{\bot }.$

Through an obvious isomorphism we may overlook $\Pi $ as being a subspace of 
$H_N$ and consider it as a space on its own right. Hence, and whenever it is
convenient, we may set $P\psi =\alpha \,,\,P\widetilde{\psi }=\widetilde{%
\alpha },$ and thus consider $P\psi \,,P\widetilde{\psi }$ as d-vectors
instead of being $N$-vectors with vanishing components in $\Pi ^{\bot }$. A
similar statement is applicable to $\Pi ^{\bot }$ and to the vectors $Q\psi
\,,Q\widetilde{\psi },$ and hence we may set $Q\psi =\beta \,,\,Q\widetilde{%
\psi }=\widetilde{\beta }.$ It is evident from (\ref{e5}) that if $\alpha =0$
then $\psi =\widetilde{\psi },$ and hence every point in the invariant
subspace $\{\tbinom 0\beta :\beta ^t\in C^{N-d}\}$ is a fixed point of the
transformation $e^{-S}.$

Let $\Psi =\{\psi _i\in H_N:i=1,.....,d\}$ be a linearly independent set of
vectors. Hence there exists at least one subspace $\Pi _K$ in which the set
of projections of these vectors is linearly independent. This last statement
is equivalent to say that the rank of the matrix $[<e_j\mid \psi
_i>],(i=1,....,d\,;\,j=1,...,N)$ is $d$. The symbol $<\,\mid \,>\,$%
designates the inner product.

We shall choose the matrix S such that

$(i)\;Q\widetilde{\psi }_i=0\,\,\,\,\,\,\,\,\,\,\,\,\,\,\,(\,i=1,.....,d)\;%
\;\;\;\;\;\;\;\;\;\;\;\;\;\;\;\;\;\;\;\;\;\;\;\;\;\;\;\;\;\;\;\;\;\;\;\;\;\;%
\;\;\;\;\;\;\,\,\,\,\;\;\;\;\;\;\;\;\;\;\;\;\;\;\;\;\;\;\;\;\;\;\;\;$

(ii) The set of vectors $P\Psi =\{P\psi _i:i=1,...,d\},$ where $P$ is the
projection corresponding to $\{1,2,..,d\},$ is linearly independent.

Requirement (ii) can always be satisfied through reordering the basis if
necessary. By (\ref{e5}), requirement (i) implies $s\alpha _i-\beta
_i=0\,(i=1,...,d),\,$or

\begin{equation}
sP\psi _i-Q\psi _i=0\,\,\,\,\,\,\,(i=1,.....,d)  \label{e7}
\end{equation}
We write (\ref{e7}) collectively as a matrix equation $s[P\Psi ]-[Q\Psi ]=0,$%
\thinspace in which $[P\Psi ]=[P\psi _1\vdots .....\vdots P\psi _d],$ $%
[Q\Psi ]=[Q\psi _1\vdots .....\vdots Q\psi _d].\;$As its columns are
linearly independent the matrix $[P\Psi ]$ is invertible, and hence 
\begin{equation}
s=[Q\Psi ][P\Psi ]^{-1}  \label{e8}
\end{equation}
Therefore requirements (i) and (ii) yield equation (\ref{e8}). It is easy to
see that equation (\ref{e8}), which embodies in it that the matrix $\left[
P\Psi \right] \;$is invertible, is fact equivalent to conditions (i) and
(ii).

With $s\,\,$so chosen, the matrix $S$ given by (\ref{e3}) has the property: $%
e^{-S}$\ projects every vector of the $d$-dimensional space $Lin\Psi $,
generated by the set of vectors $\Psi =\{\psi _{1,.....,}\psi _d\}$, onto
the model space $Lin\{\alpha _1,....,\alpha _d\}\equiv \Pi .$\thinspace This
follows immediately from requirement (i) and linearity of $e^{-S}.$ In other
words, an arbitrary vector $\tbinom \alpha {s\alpha }\in Lin\Psi $ is mapped
by $e^{-S}\;$to $\tbinom \alpha 0\in \Pi $. The operator $e^{-S}\,$is not a
projection operator as implied by the mathematical definition of a
projection operator. The word ''project'' however is used here in a
geometrical sense to describe an operation in which every vector of a
certain subspace (visualized as hyperplane) is mapped to a vector that has
the same first $d\;$components, whereas its remaining components are zeros
(visualized as a vector in a coordinate hyperplane). Also if $\phi \notin
Lin\Psi ,$ then its image $\widetilde{\phi }$ is not in the model space. The
proof of the last fact relies on the regularity of $e^{-S}$, which implies
that the image of the independent set $\{\psi _1,....,\psi _d,\phi \}$
namely $\{\alpha _1,....,\alpha _d,\widetilde{\phi }\}$ is linearly
independent, and hence $\phi \notin Lin\{\alpha _1,....,\alpha _d\}=\Pi $ .
The vector $\widetilde{\phi }\,$therefore has at least one non-vanishing
component outside the space $\Pi $ .

The operator $e^{-S}$, with $s$ given by (\ref{e8}), as projects the
subspace $Lin\Psi $ orthogonally on $\Pi $, is thus determined solely by $%
Lin\Psi $ and $\Pi $ , and is independent of the particular choice of a set
of $d$ independent vectors in $Lin\Psi $ . Indeed if $\Psi ^{^{\prime
}}=\{\psi _1^{^{\prime }},...,\psi _d^{^{\prime }}\}$ is another set of
independent vectors in $Lin\Psi $, then 
\begin{equation}
\psi _i^{^{\prime }}=\stackunder{j=1}{\stackrel{d}{\sum }}c_{ji}\psi
_j\,\,\,\,\,(i=1,...,d)  \label{e9}
\end{equation}
where $c_{ji}\;$are constants. Denoting the matrix whose elements are $%
c_{ji}(i,j=1,....,d)$ by $C$, and the matrices whose columns are $\psi _i$
and $\psi _i^{^{\prime }}\,$by $[\Psi ]$ and $[\Psi ^{^{\prime }}]$
respectively, we write the last relation as $[\Psi ^{^{\prime }}]=[\Psi ]C$.
Equivalently we have $[P\Psi ^{^{\prime }}]=[P\Psi ]C$ and $[Q\Psi
^{^{\prime }}]=[Q\Psi ]C$. Substituting from these equations for $[P\Psi ]$
and $[Q\Psi ]$ in (\ref{e8}) we get $s=[Q\Psi ^{^{\prime }}][P\Psi
^{^{\prime }}]^{-1},$ which proves our assertion.

We finally note that as $e^{-S}$ is invertible, the inverse image of every
vector $\alpha \in \Pi $, which also is identifiable with\thinspace $\tbinom
\alpha 0$, \ is retrievable as\thinspace $\binom \alpha {s\alpha }.$

\section{Lee and Suzuki Transformation}

Let $O$ be a Hermitian $N\times N$ matrix, with an independent set of
eigenvectors$\{\psi _i:i=1,....,N\}$, and consider the eigenequation 
\begin{equation}
O\,\psi _i=E_i\psi _i\,\,\,\,\,\,\,(i=1,.....,N)  \label{e10}
\end{equation}
Applying the Lee and Suzuki similarity transformation \cite{Lee} to the
matrix $O$ and to the truncated space $H_N$, we obtain 
\begin{equation}
\widetilde{O}\,\widetilde{\psi }_i=E_i\widetilde{\psi }_i\;\;\;\;%
\;(i=1,....,N)\text{,}  \label{e11}
\end{equation}
where we have used tilde to designate transformed quantities so that 
\begin{equation}
\widetilde{O}=e^{-S}O\,e^S,\,\,\,\,\,\,\,\,\,\,\widetilde{\psi }=e^{-S}\psi .
\label{e12}
\end{equation}
Our work will be distinguished from that of Lee and Suzuki through our
identification of additional freedom in the choice of S. Multiplying both
sides of (\ref{e11}) by$\;P\;$and injecting $\;I=P+Q$ \ conveniently in the
right hand side we get 
\begin{equation}
P\widetilde{O}P\widetilde{\psi }_i+P\widetilde{O}Q\widetilde{\psi }_i=E_iP%
\widetilde{\psi }_i\,\,\,\,\,\,\,\,\,(i=1,....,N).  \label{e13}
\end{equation}
In a similar way we get 
\begin{equation}
Q\widetilde{O}P\widetilde{\psi }_i+Q\widetilde{O}Q\widetilde{\psi }_i=E_iQ%
\widetilde{\psi }_i\,\,\,\,\,\,\,(i=1,....,N).  \label{e14}
\end{equation}
We shall choose the transformation (\ref{e12}) such that there exists a
subset $J\subset \{1,...,N\}$ with $card\,J=d,\;$for which (i) the set of
vectors $\{P\widetilde{\psi }_i:i\in J\}\,$is linearly independent, and (ii) 
$Q\widetilde{\psi }_i=0\,(i\in J).$ Such a choice, as we have seen in the
previous section, is certainly possible.\ 

Proposition 1: Let $J\subset \{1,...,N\}\;$be such that the set\ $\{P\psi
_i:i\in J\}$ is linearly independent. The following assertions concerning
the Lee and Suzuki transformation are equivalent:

A1. $Q\widetilde{\psi }_i=0\,(i\in J)$

A2. $s_J=[Q\Psi _J][P\Psi _J]^{-1}$\ 

A3. (i) the decoupling equation $Q\widetilde{O}P=0$ holds, and

\ \ (ii) $P\psi _i\;(i\in J)$ are eigenvectors of$\;P\widetilde{O}P.$\ 

Proof. We have seen in section 2 that the assertions A1 and A2 are
equivalent (this expression of $s_J$ was first given by Navratil and Barrett 
\cite{Navratil} ) . To prove that assertion A1 implies A3, we set $Q%
\widetilde{\psi }_i=0\;(i\in J)$\ in (\ref{e13}) and (\ref{e14}) to find
that $P\widetilde{\psi }_i$ $(i\in J)$ are eigenvectors of $P\widetilde{O}P,$
and $Q\widetilde{O}P\widetilde{\psi }_i=0$\ $(i\in J).$ Due to the linear
independence of \ $P\psi _i\;(i\in J),$ the later d equations\ imply that $Q%
\widetilde{O}P=0$\ . Conversely, if \ $\alpha _k\;(i=1,...,d)$\ are linearly
independent eigenvectors of $\;P\widetilde{O}P$\ then the N-vectors $\tbinom{%
\alpha _k}0(k=1,...,d)$ are eigenvectors of $\widetilde{O.}$\ It follows
that the inverse image of these vectors$\;\{e^S\binom{\alpha _k}%
0:k=1,...,d\} $\ coincides with a subset $\Psi _J=\{\psi _i:i\in J\}$\ of
eigenvectors of $O.$ The subset $\Psi _J$ clearly fulfills assertion 1.
Hence A1 is equivalent to A3.

\ \ \ \ \ \ \ \ \ \ \ \ \ \ \ \ \ \ \ \ \ \ \ \ \ \ \ \ \ \ \ \ \ \ \ \ \ \
\ \ \ 

\section{The Effective Form}

When the transformed operator $\widetilde{O}$ is such that $Q_K\widetilde{O}%
P_K=0,$ for some subset $K\subset \{1,.....,N\},$ with $card\,K=d,$ we refer
to the operator $O_{eff}\equiv P_K\widetilde{O}P_K$ as an effective
representative of the operator $O\,$\thinspace corresponding to the model
space $\Pi _K,$ and to the form taken by $\widetilde{O}$ as an effective
form. When $\widetilde{O}$ is in an effective form corresponding to the
model space $\Pi ,$ the matrix elements $(O_{eff})_{ij}$ are all zero except
those for which $i,j\leq d$, and consequently we make the identification $%
O_{eff}:\Pi \rightarrow \Pi ,$ in which $O_{eff}$ is considered a $d\times d$
matrix. In a similar way we treat $Q\widetilde{O}Q$ as an $(N-d)\times (N-d)$
matrix.

We elaborate here on the effective form and develop a more explicit
framework. We write the eigenequation (\ref{e10}) as 
\begin{equation}
\left( 
\begin{array}{cc}
a & b \\ 
b^{+} & f
\end{array}
\right) \left( 
\begin{array}{c}
\alpha _i \\ 
\beta _i
\end{array}
\right) =E_i\left( 
\begin{array}{c}
\alpha _i \\ 
\beta _i
\end{array}
\right) \,\,\,\,\,\,\,\,\,\,\,(i=1,.....,N)\text{ }  \label{e16}
\end{equation}
where the matrix $O$ has been partitioned to submatrices corresponding to $%
\Pi $ and $\Pi ^{\bot },$ with $a$ is a $d\times d$ matrix. By (\ref{e12})
the last equation is transformed to

\begin{equation}
\left( 
\begin{array}{cc}
a+bs & b \\ 
-s(a+bs)+b^{+}+fs & f-sb
\end{array}
\right) \left( 
\begin{array}{c}
\alpha _i \\ 
\beta _i-s\alpha _i
\end{array}
\right) =E_i\left( 
\begin{array}{c}
\alpha _i \\ 
\beta _i-s\alpha _i
\end{array}
\right)  \label{e17}
\end{equation}
The later equation is equivalent to (\ref{e13}) and (\ref{e14}) together. It
is easy to check that every $s_J$ , given as in proposition 1, puts $%
\widetilde{O}$ into an effective form corresponding to some model space $\Pi 
$. In other words, every $s_J$\thinspace is a solution to the decoupling
equation 
\begin{equation}
Q\widetilde{O}P\equiv -s(a+bs)+b^{+}+fs=0.  \label{e18}
\end{equation}
To demonstrate the converse we assume that the later equation is satisfied
by some $s$, and hence the action of $Q\widetilde{O}P$ on any vector in $\Pi 
$ is zero. In particular this action is zero for all vectors $\alpha _i$
such that $\psi _i=\tbinom{\alpha _i}{\beta _i},(i=1,....,N)$ are
eigenvectors of $O$, and hence 
\begin{equation}
-s(a+bs)\alpha _i+b^{+}\alpha _i+fs\alpha _i=0\,\,\,\,\,\,(i=1,....,N)
\label{e19}
\end{equation}
Making use of (\ref{e16}) we reduce the last equation to the eigenequation 
\begin{equation}
(f-sb)(s\alpha _i-\beta _i)=E_i(s\alpha _i-\beta _i)\,\,\,\,\,\,(i=1,....,N)
\label{e20}
\end{equation}
which is the same as embodied in equation (\ref{e17}) but now extended to
all $i$. However, not all vectors $s\alpha _i-\beta _i$ can be eigenvectors
of $(f-sb)$ because the later operator has only $N-d$ eigenvectors. It
follows that there exists a subset $J$ consisting of $d$ elements of $%
\{1,....,N\}$ such that $s\alpha _i-\beta _i=0\,(i\in J),$which implies that 
$s=s_J,$ as given in proposition 1.

We list here the following comments on the effective form assuming from now
on that $\widetilde{O}$ is in such a form. i.e. the transformation (\ref{e11}%
) is such that $Q\widetilde{O}P=0$.

1. If $\widetilde{O}$ is the effective form corresponding to the model space 
$\Pi $ then the right hand-side of the secular (characteristic) equation $%
\det (O-EI_N)=0\,$can be factorized to a product of two polynomials; one of
which is of degree $d$ in $E$ 
\begin{equation}
\det (O_{eff}-EI_d).\det (Q\widetilde{O}Q-EI_{N-d})=0.  \label{e21}
\end{equation}
The eigenvalues of $O$ is the set of zeros of these two polynomials. In
practical problems the secular equation of $O_{eff}$ can be solved
numerically as it is of \ low degree in $E$, whereas that of $Q\widetilde{O}%
Q $ is of \ high degree in $E$ and it is often hopeless to approach it for
direct solution. One may apply the method of effective form described in the
previous section afresh to the operator $Q\widetilde{O}Q$. Or alternatively
one may pick up a new set of eigenvectors, say $\Psi _{J^{\prime }},$%
determine $s_{J^{\prime }}$, and consequently a new effective form.
Alternatively the matrix $s\;$could be determined by iterative methods \cite
{Kuo,Cloiz,zheng}.

2. If $P\psi _i$ is an eigenvector of $O_{eff}$ corresponding to $E_i$ then
by (\ref{e13}) $P\widetilde{O}Q\widetilde{\psi }_i=0$ which implies that the 
$(N-d)$-vector $Q\widetilde{\psi }_i$ is complex orthogonal to the rows of $%
d\times (N-d)$ matrix $P\widetilde{O}Q,$ and the vector $Q\widetilde{\psi }%
_i $ is not necessarily zero. Therefore, if $\alpha $ is an eigenvector of\ $%
O_{eff}$ then, though $\tbinom \alpha 0$ is an eigenvector of $\widetilde{O}$
belonging to the eigenvalue $E_i,$ there may exist another eigenvector $%
\tbinom \alpha \gamma $\ of $\widetilde{O}$ that belongs to the same
eigenvalue $E_i.$ In the latter case $b\gamma =0$ and $\gamma $ is an
eigenvector of $Q\widetilde{O}Q$ belonging to the eigenvalue $E_i$ . It is
clear that $\tbinom 0\gamma $ is an eigenvector of $\widetilde{O}$ that
belongs to the eigenvalue $E_i$. We summarize the latter observations by the
following proposition

Proposition 2. Let $\phi $ be an eigenvector of $\widetilde{O}$ belonging to
the eigenvalue $E$.

(i) If $Q\phi =0$ then $E$ is an eigenvalue of $O_{eff}$ to which the
eigenvector $P\phi $ belongs.

(ii) If $Q\phi \neq 0$ then $Q\phi $ is an eigenvector of $Q\widetilde{O}Q$
belonging to the eigenvalue $E.$ If in addition, $P\phi \neq 0,$ then $P\phi 
$ is not an eigenvector of $O_{eff}$ unless $bQ\phi =0.$ In the latter case $%
E$ is a common eigenvalue of $O_{eff}$ and $Q\widetilde{O}Q$ to which the
independent eigenvectors $\tbinom{P\phi }0$ and $\tbinom 0{Q\phi }$ belong.
In the latter case the spectra of $O_{eff}$ and $Q\widetilde{O}Q$ intersect.

(iii) $P\phi $ is an eigenvector of $O_{eff}$ does not necessitate that $%
Q\phi =0.$ However if the spectra of $O_{eff\text{ }}$and $Q\widetilde{O}Q$
do not intersect in $E,$ then $Q\phi =0\Leftrightarrow P\phi $ is an
eigenvector of $O_{eff}$ belonging to the eigenvalue $E.$

3. If the matrix $\left[ P\Psi _J\right] $ \thinspace is singular for some
choice of model space, say $\Pi ,$ then we have to replace it by another $%
\Pi ^{\prime }$ such that the matrix $\left[ P^{\prime }\Psi _J\right] $ is
invertible. There certainly exists such a new choice of model space,
otherwise the set $\Psi _J$ would be linearly dependent.

We demonstrate here that for a given set of eigenvectors $\Psi _J$\thinspace
, two legitimate choices of model spaces lead to two effective
representatives which are related by a similarity transformation. Let $\Pi $
and $\Pi ^{\prime }$ be two legitimate choices and denote the projections on
the corresponding model spaces by\ $P\;$and $\;P^{\prime }$ respectively.
This leads to two distinct $s$, say $s\;$and $s^{\prime }$ , and hence to
two distinct effective representatives $O_{eff}=P\widetilde{O}P$ and $%
O_{eff}^{^{\prime }}=P^{\prime }\widetilde{O}^{\prime }P^{\prime }.$ If $%
\{E_i:i\in J\}$ is the set of eigenvalues to which $\Psi _J$ belong, then

\begin{equation}
O_{eff\,}[P\Psi _J]=[P\Psi _J]\,\Lambda _J\,,\,O_{eff}^{\prime }\,[P^{\prime
}\Psi _J]=[P^{\prime }\Psi _j]\,\Lambda _J  \label{e22}
\end{equation}
where $\Lambda _J$ is a diagonal matrix with diagonal elements $(E_i:i\in
J). $ From (\ref{e22}) we deduce that 
\begin{equation}
O_{eff}\,=[P\Psi _J][P^{\prime }\Psi _J]^{-1}O_{eff\,}^{\prime }[P^{\prime
}\Psi _J][P\Psi _J]^{-1}  \label{e23}
\end{equation}
which proves our claim.

Each independent set of eigenvectors $\Psi _{J\text{ \thinspace }}$provide
at least one model space $\Pi _K.$ The number of possible choices of $\Pi
_K\,$is not less than one and not greater than\thinspace $\tbinom Nd,$%
\thinspace which is of course the number of independent sets of projections $%
\{P_K\Psi _J:cardK=d,\,K\subset \{1,....,N\;\}\}$ . All such choices lead of
course to the same set of eigenvalues $\Lambda _J.$

If the eigenvalues of $O$ are non-degenerate then different choices of $\Psi
_J$ out of the set of $N$ independent eigenvectors $\Psi ,$ result in
effective representatives with different spectra. The total choices of
inequivalent effective representatives corresponding to $O$ is $\tbinom
Nd;\; $and within each of these there are a maximum number of $\tbinom Nd$
equivalent representatives.

The above-identified freedoms are new and extend the work of Lee and Suzuki.

\section{Spectral Representation of $O_{eff}$}

Let $O_{eff}$ be an effective representative of the operator $O$ in the
model space $\Pi $, and let $\{E_i:i\in J\}$ be the spectrum of $%
\,\,\,\,O_{eff}$, to which the vectors $P\Psi _i(i\in J)$ belong, so that $%
O_{eff}P\psi _i=E_iP\psi _i\,(i\in J).$ Since each $P\psi _i$ lies in the
model space we have 
\begin{equation}
P\psi _i=\stackunder{\mu =1}{\stackrel{d}{\sum }}c_{i\mu }\,e_\mu
\;\;\;\;(i\in J)  \label{e24}
\end{equation}
\begin{equation}
<P\psi _i\mid P\psi _j>=\stackunder{\mu =1}{\stackrel{d}{\sum }}c_{i\mu
}^{*}\;c_{j\mu }\equiv \gamma _{ij}\,\,\,\,(i,j\in J)\text{ }  \label{e25}
\end{equation}
The matrix $\gamma $ is clearly Hermitian, and determines the overlap the
eigenvector of $O_{eff}$ one with respect to another. Let 
\begin{equation}
\chi =\stackunder{j\in J}{\stackrel{}{\sum }}b_jP\psi _j  \label{e26}
\end{equation}
be an arbitrary vector in the model space, then 
\begin{equation}
<P\psi _i\mid \chi >=\stackunder{j\in J}{\stackrel{}{\sum }}\gamma
_{ij}\,b_j\,\,\,\,\,\,\,\,\,\,(i\in J)  \label{e27}
\end{equation}
Hence 
\begin{equation}
\stackrel{}{\stackunder{i\in J}{\sum }}\,\gamma _{ki}^{-1}<P\psi _i\mid \chi
>=b_k\,\,\,\,\,\,\,\,\,(k\in J)  \label{e28}
\end{equation}
where $\gamma ^{-1\;}$is the inverse of the matrix $\gamma $. It is clear
that $\gamma ^{-1}\;$always exists since $P\psi _i\,(i\in J)$ are linearly
independent. Applying $O_{eff}\;\,$to$\,\chi $ where $b_k$ are given by (\ref
{e28}\thinspace )\thinspace \ we get 
\begin{equation}
O_{eff}\mid \chi >=\stackunder{i,k\in J}{\sum }\gamma _{ki}^{-1}<P\psi
_i\mid \chi >E_k\;P\psi _k  \label{e29}
\end{equation}
This yields

\[
O_{eff}=\stackunder{i,k\in J}{\sum }E_i\gamma _{ik}^{-1}\mid P\psi _i><P\psi
_k\mid 
\]
which expresses $O_{eff}\;$in terms of quantities pertaining to the model
space.

\section{A Complete Set of Effective Representatives}

Let $O_1\equiv H$ be the Hamiltonian of a physical system and $O^2,.....,O^c$
be a set of observables pertaining to the system so that the set of
observables\thinspace $\Gamma =\{O^1,O^2,....,O^c\}$ is complete. It follows
from the latter assumption that 
\begin{equation}
\lbrack O^\rho ,O^\sigma ]=0\,\,\,\,\,(\rho ,\sigma =1,....,c)  \label{e31}
\end{equation}
The energy eigenvectors $\{e_i:i=1,2,....\}$ of any suitably chosen simple
Hamiltonian could be taken as a basis for the Hilbert space of wave
functions of the physical system. For example, these could be the
eigenstates of the simple harmonic oscillator, when considering the bound
states of the nucleus. Obeservables pertaining to the system are represented
by Hermitian matrices in terms of this basis. Unless the matrices
representing observables are given by recurrence formulae, we have to be
content with finite matrix approximations, which imply truncating the
infinite basis $\{e_i\}_1^\infty \,$at some sufficiently large term $N$. The
space generated by the truncated basis$\,\,[e_1,....,e_N]\equiv H_N\,$\ $\,$%
will hopefully contain good approximations of all states of interest to the
problem we consider.

It must be noted that, whenever the eigenvalue problem is to be solved
numerically, which is usually the case in physically interesting problems,
truncation is an inevitable task. We note that truncating an infinite basis
by a finite one with a sufficiently large number of basis elements is
justified by the fact that the sequence $(e_N)$ tends weakly to zero as $N$
tends to infinity. This means that for every wave function $\psi \in H$%
\thinspace \thinspace the sequence of numbers $(<e_N\mid \psi >)\,$%
\thinspace tends to zero as $N$ tends to $\inf $inity \cite{Fano,Prugovecki}%
. Alternatively, an upper cutoff, $N,\;$can be safely applied without
seriously changing the low-lying properties.

It is noted that all the algebra carried out in the previous section, or to
be carried out in the forthcoming discussion, is valid for infinite matrices
as much as it is valid for finite ones, and hence we may replace N by $%
\infty $ without affecting the validity of these results.

The Hermitian commuting set of matrices $\Gamma $ is complete, and there
exists accordingly a complete set of simultaneous eigenfunctions $\psi _i$
of the observables $O^\sigma $ such that

\begin{equation}
O^\sigma \psi _i=E_i^\sigma \psi _i\;\;\;\;\;(i=1,...,N;\sigma =1,....,c)
\label{e32}
\end{equation}
where $E_i^\sigma $ are the eigenvalues of the observable $O^\sigma $ to
which the eigenvector $\psi _i$ \thinspace \thinspace \thinspace \thinspace
belongs. The eigenvectors given by (\ref{e32}) are preserved when the
similarity transformation (\ref{e12} ) is applied to the Hilbert space of
wave functions $H_N\;$and to the operators acting on $H_N$, and hence

\begin{equation}
\widetilde{O}^\sigma \widetilde{\psi }_i=E_i^\sigma \widetilde{\psi }%
_i\,\,\,\,(i=1,...,N\,;\sigma =1,...,c)  \label{e33}
\end{equation}
Assume that the eigenvectors $\widetilde{\psi }_1,....,\widetilde{\psi }_d$
\thinspace \thinspace are such that the set $\{P\widetilde{\psi }_1,.....,P%
\widetilde{\psi }_d\}$ is linearly independent, and take 
\begin{equation}
s=[Q\widetilde{\psi }_1\left. .......\right. Q\widetilde{\psi }_d][P%
\widetilde{\psi }_1\left. ......\right. P\widetilde{\psi }_d]^{-1}
\label{e34}
\end{equation}
The matrix $s$ is the same for observables forming the complete set $\Gamma $%
, for it is constructed of the same subset of the simultaneous eigenvectors
of $\widetilde{O}^\sigma (\sigma =1,...,c)$. The resulting transformed
observables $\widetilde{O}^\sigma $, have the same effective form, and hence
have $P\widetilde{\psi }_i\,(i=1,...,d)$ as a common subset of eigenvectors $%
\{\widetilde{\psi }_i:i=1,...,N\}$. Define a set of effective
representatives 
\begin{equation}
O_{eff}^\sigma =P\widetilde{O}^\sigma P\;\;\;\;\;(\sigma =1,....,c)
\label{e35}
\end{equation}
and hence

\thinspace \thinspace \thinspace 
\begin{equation}
\,O_{eff}^\sigma \,P\widetilde{\psi }_i=E_i^\sigma P\widetilde{\psi }%
_i\,\,\,\,\,\,\,\,\,\,\,\,(i=1,....,d;\sigma =1,....,c)  \label{e36}
\end{equation}
It follows, and since the set $\{P\psi _i:i=1,...,d\}$\thinspace is complete
in the model space $\Pi $, that 
\begin{equation}
\lbrack O_{eff}^\rho \,,O_{eff}^\sigma ]=0\,\,\,\,\,\,\,\,\,\,\,(\sigma
,\rho =1,...,c)  \label{e37}
\end{equation}

The effective Hamiltonian $H_{eff}\equiv O_{eff}^1$ and the effective
representatives $O_{eff}^\sigma (\sigma =2,...,c)\,$we have constructed have
the virtue that the symmetries exhibited by the original Hamiltonian $H\,$%
are carried over to$\,H_{eff}\,\,$with the effective representatives $%
O_{eff}^\sigma \;(\sigma =2,...,c)$ playing the role of generators of
symmetry for $H_{eff}.$

The matrices (\ref{e35} ) are obviously non-Hermitian and, consequently, the
expectation value of an effective representative in a state $P\psi $ in the
model space is generally a complex number. An exception to this fact is that
when $P\psi $ is an eigenvector $P\psi _i\;$of $O_{eff}$ . In this case 
\begin{equation}
<O_{eff}>_{P\psi _i}=<P\psi _i\mid O_{eff}\mid P\psi _i>/\parallel P\psi
_i\parallel ^2=E_i  \label{e38}
\end{equation}

\section{ A Second Type of Effective Representative.}

The role of an effective operator seems limited to producing some of the
eigenvalues of the original observable. However, we may enhance the scheme
of ''effectiveness'' and make a further step as follows: The matrix $S$
which is determined by iterative methods \cite{Kuo,zheng,Jvary} and utilized
to construct the effective representative $O_{eff}$ can also be utilized to
construct an effective representative of a second type $\overline{O}_{eff}\;$%
that satisfy the property 
\begin{equation}
<\psi \left| O\right| \phi >=<P\psi \mid \overline{O}_{eff}\mid P\phi >
\label{e39}
\end{equation}
for all $\psi ,\phi \in Lin\{\psi _1,....,\psi _d\}$ . Using (\ref{e12}) and
the definition of the adjoint operator, we have 
\begin{eqnarray}
\ \left\langle \psi \left| O\right| \phi \right\rangle &=&\left\langle e^{-S}%
\widetilde{\psi }\mid O\mid e^{-S}\widetilde{\phi }\right\rangle
=\left\langle \widetilde{\psi }\mid e^{-S^{+}}O\;e^{-S}\mid \widetilde{\phi }%
\right\rangle  \label{e40} \\
\ &=&\left\langle P\widetilde{\psi }\mid e^{-S^{+}}O\,e^{-S}\mid P\widetilde{%
\phi }\right\rangle =\left\langle P\psi \mid Pe^{-S^{+}}O\,e^{-S}P\mid P\phi
\right\rangle  \nonumber
\end{eqnarray}
The requirement ( \ref{e39}) is fulfilled on taking

\begin{equation}
\overline{O}_{eff}=Pe^{-S^{+}}O\,e^{-S}P  \label{e41}
\end{equation}
In particular $<\psi _i\mid O\mid \psi _j>=<\alpha _i\left| \overline{O}%
_{eff}\right| \alpha _j>.\,$

We therefore associate with every observable $O^\sigma \in \Gamma \,$two
effective representatives. The first, $O_{eff}$, serves to determine some of
the eigenvalues of $O^\sigma $ and the projection of the corresponding
eigenvectors on the model space; the second $\overline{O}_{eff}^\sigma $ has
the important property : the matrix elements of the original operator $%
O^\sigma $ with respect to any basis in the space $Lin\{\psi _{1,....,}\psi
_d\}$ is given in terms of $\overline{O}_{eff}^\sigma $ and the projected
basis in the model space. It is evident that the last matrix can be
calculated easily since $\overline{O}_{eff}^\sigma $ is known whenever $S$
is known, and since the basis elements of the model space have finite
components. We mention that the matrix $<P\psi _i\mid \overline{O}_{eff}\mid
P\psi _j>$ is not the matrix of $\overline{O}_{eff}$ since $\{P\psi
_i\}_{i=1}^d$ is not orthogonal. In particular, and if$\;\psi \in Lin\Psi $\
then 
\[
<O>_\psi =<P\psi \mid \overline{O}_{eff}\mid P\psi >=\left\| P\psi \right\|
^2<\overline{O}_{eff}>_{P\psi } 
\]
Expressed in words, the expectation value of the observable $O$ in the state 
$\psi \in Lin\Psi $ is equal to the expectation value of its representative
of the second type in the projection of the given state on the model space
times the square norm of this projection.

For observables $O$ that do not commute with all elements of the complete
set $\Gamma \,$ we can define only one effective representative, that is the
effective representative of the second type $\overline{O}_{eff}.$\thinspace
This serves to give a portion of the transition matrix of $O$, namely that
which corresponds to a basis of $Lin\Psi .$

A systematic study of the system is achieved by decomposing the space $H_N$
into linear subspaces $Lin\Psi _{J_r}\,(r=1,2,...,a),\,$with $J_r\cap
J_s=\emptyset $ if $(r\neq s),\,\;$so that

\begin{equation}
H_N=Lin\Psi _{J_1}\oplus \,....\,\oplus \,Lin\Psi _{J_a}  \label{e42}
\end{equation}
Now in each subspace $Lin\Psi _{J_r}$ we assign to every observable $O$ in a
complete set of observables an effective representative of the first type $%
O_{r\,eff}$ and an effective observable of the second type $\overline{O}%
_{r\,eff}$. These effective representatives, of first or second type, differ
from one subspace to another as does $s_J.$ If representatives of the first
type are all obtained, all eigenvalues of the full observable $O$ become
known. Also if $\chi $, $\chi ^{\prime }\in Lin\Psi _{J_r}$ then we have $%
<\chi \left| O\right| \chi ^{\prime }>=<P_r\chi \left| \overline{O}%
_{r\;eff}\right| P_r\chi ^{\prime }>$ where$\;P_r$ denotes the projection on
the model space corresponding to the subspace $Lin\Psi _{J_r}$. Although
similar relations are valid for every two vectors in the same subspace, one
can not express $<\chi \left| O\right| \chi ^{\prime }>$ in terms of
representatives of second type when $\chi $ and $\chi ^{\prime }$ belong to
different subspaces, and consequently when they are arbitrary vectors in $%
H_N $.

\section{Towards Practical Applications}

Following the traditional lines of thinking for many-body problems, we
suggest that $S$ is developed for small sub-systems and used as an
approximation for the full $S.$ For example two and three body problems may
be solved with high precision using current numerical techniques \cite
{Navratil,Jvary}.\ A set of solutions $P\psi _i(i=1,...,d)$ is selected, $S$
is evaluated and the resulting $H_{eff}$ is then used in many body problems
within the appropriately restricted model space. Detailed tests will be
needed for specific Hamiltonians to determine the efficacy of this approach
and the utility of the various freedoms we have identified within the
present work.

\section{Acknowledgments}

The authors thank Kenji Suzuki for a valuable discussion. Cesar Viazminsky
acknowledges the generous financial support provided by the University of
Aleppo in Syria. James P Vary acknowledges partial support by the U.S.
Department of Energy under grant No. DE-FG02-87ER-40371, Division of High
Energy and Nuclear Physics.

\end{document}